\documentclass{appolb}
\usepackage{epsfig}

\begin{document}
\pagestyle{plain}
\newcount\eLiNe\eLiNe=\inputlineno\advance\eLiNe by -1
\title{Searches for New Physics by the H1 Experiment at HERA%
\thanks{Presented at the PIC 2007 conference, on behalf of the H1 Collaboration}%
}
\author{trinh thi nguyet
\vspace{-0.45cm}
\address{Centre de Physique des Particules de Marseille, Marseille,
13288 cedex 09, France}}
\vspace{-0.75cm}
\maketitle
\vspace{-0.75cm}
\begin{abstract}
The high energy program of the HERA {\it ep} collider ended in March 2007, where data equivalent to an integrated luminosity of $\sim$ $0.5$~fb$^{-1}$
has been collected by the H1 experiment. In this context, some of the most recent results from H1 about searches
for new phenomena are presented.
\end{abstract}
\begin{figure}[htbp] 
  \begin{center}
    \includegraphics[width=4.5cm]{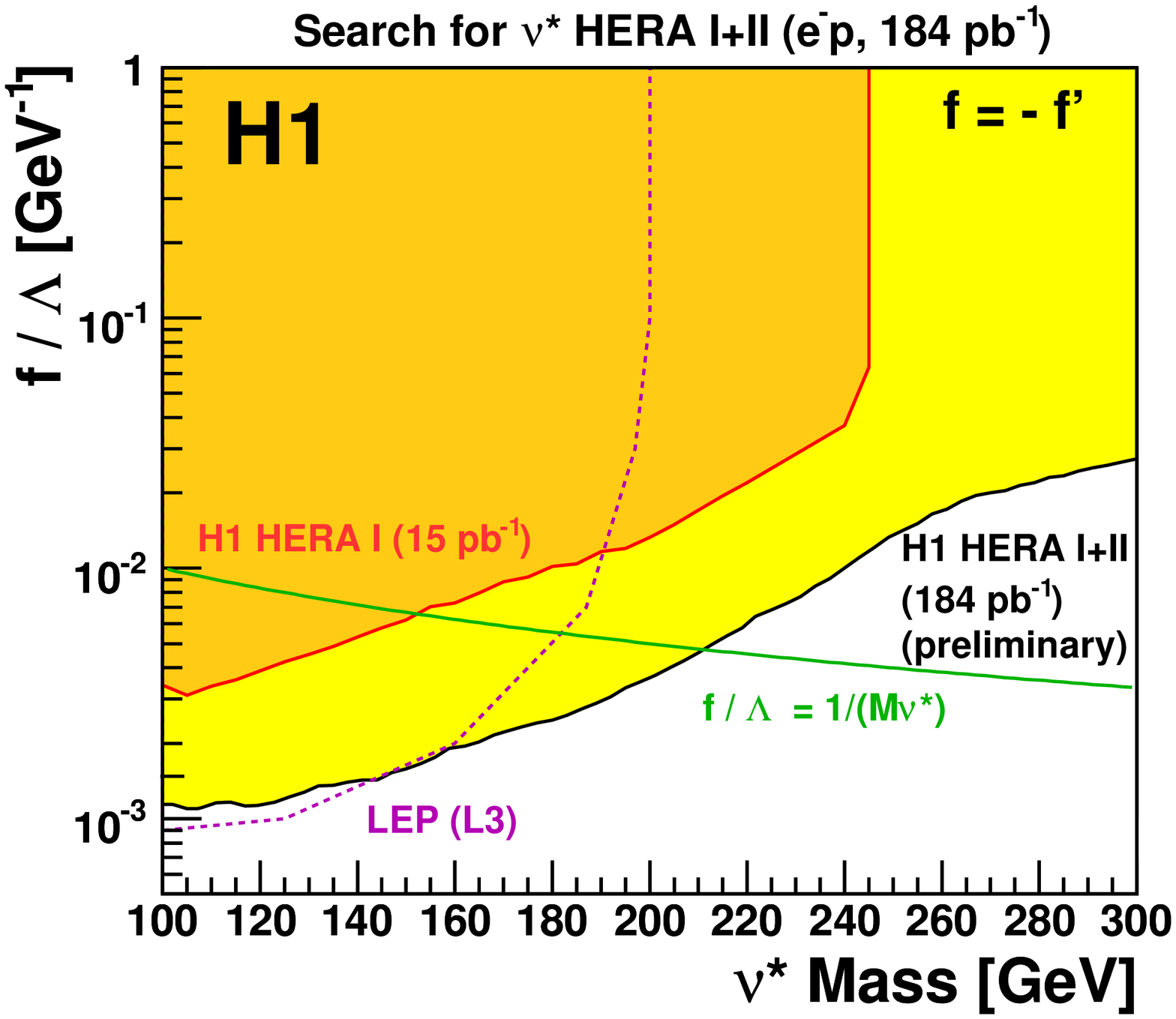}
    \includegraphics[width=4.5cm]{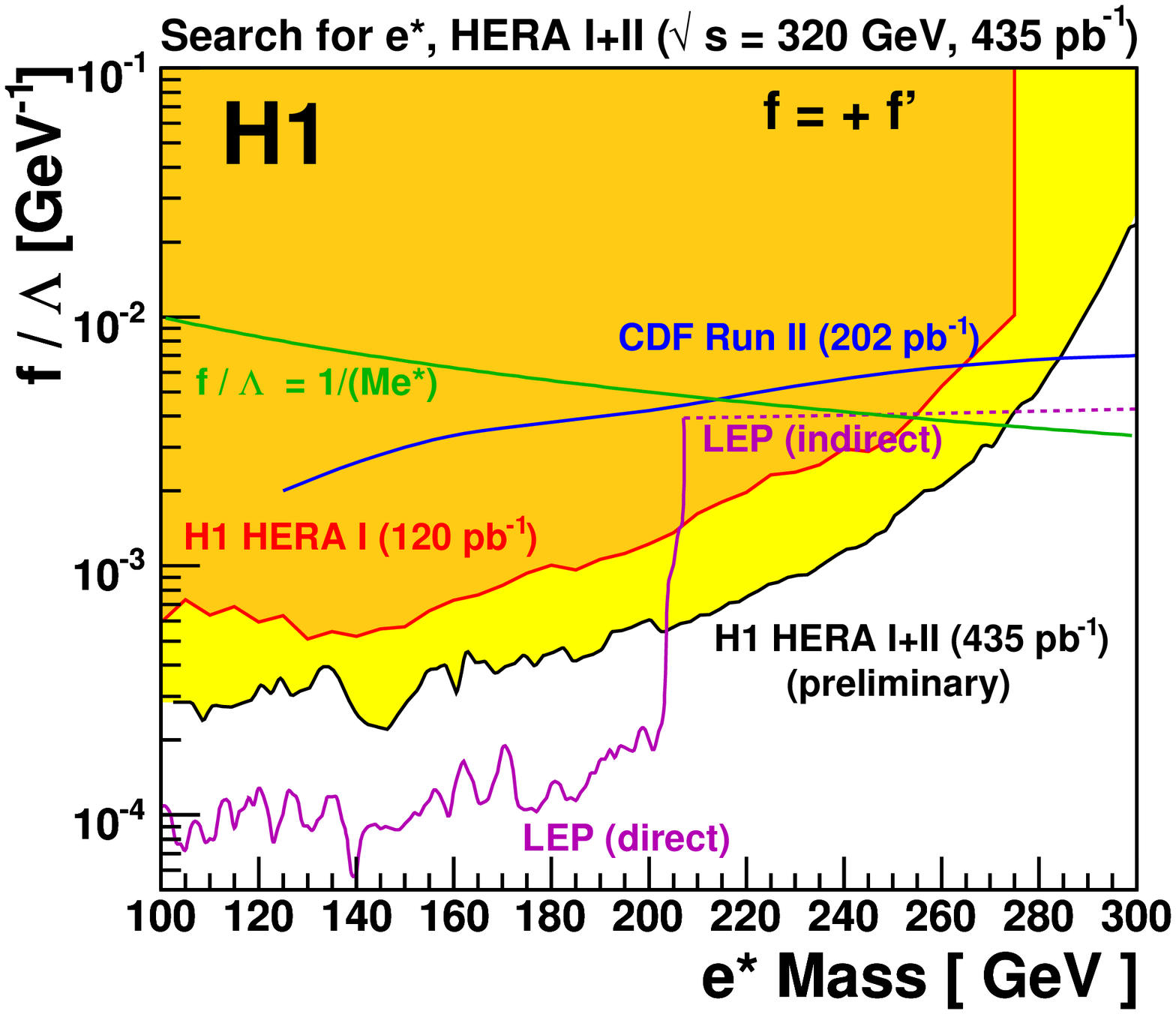}
  \end{center}
  \vspace{-0.75cm}
  \caption{Limits at 95\% CL on the coupling constants $f/\Lambda$ for excited neutrino (left) and electron (right).}
\label{fig:LimitCouplingNustar}  
\end{figure}

The direct observation of excited states of fermions ($f^{*}$) as a natural consequence of the compositeness models~\cite{Harari, Boudjema}  via their decay into a fermion and a gauge boson would be an evidence for a new level of substructure. New H1 results about the search for excited leptons ($e^{*}$,~$\nu^{*}$) make use of a total luminosity of up to $435$~pb$^{-1}$. In the absence of a signal, $95\%$~CL upper limits on the coupling $f/\Lambda$ are derived as a function of the excited lepton mass as presented in figure~\ref{fig:LimitCouplingNustar}. The new limits extend at high masses previous bounds reached at LEP and Tevatron colliders.           

\begin{figure}[htbp] 
  \begin{center}
    \includegraphics[width=4.5cm]{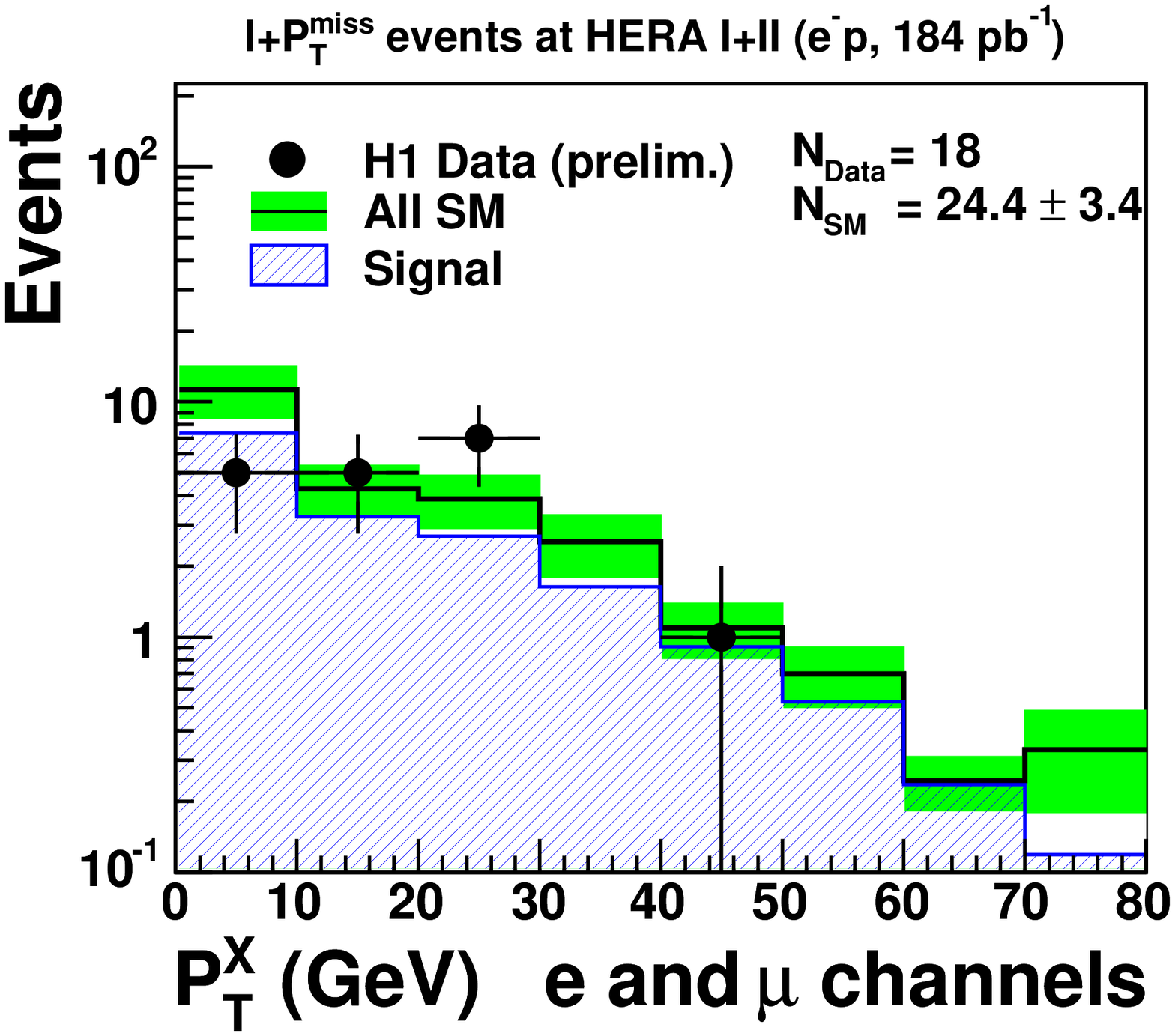}
    \includegraphics[width=4.5cm]{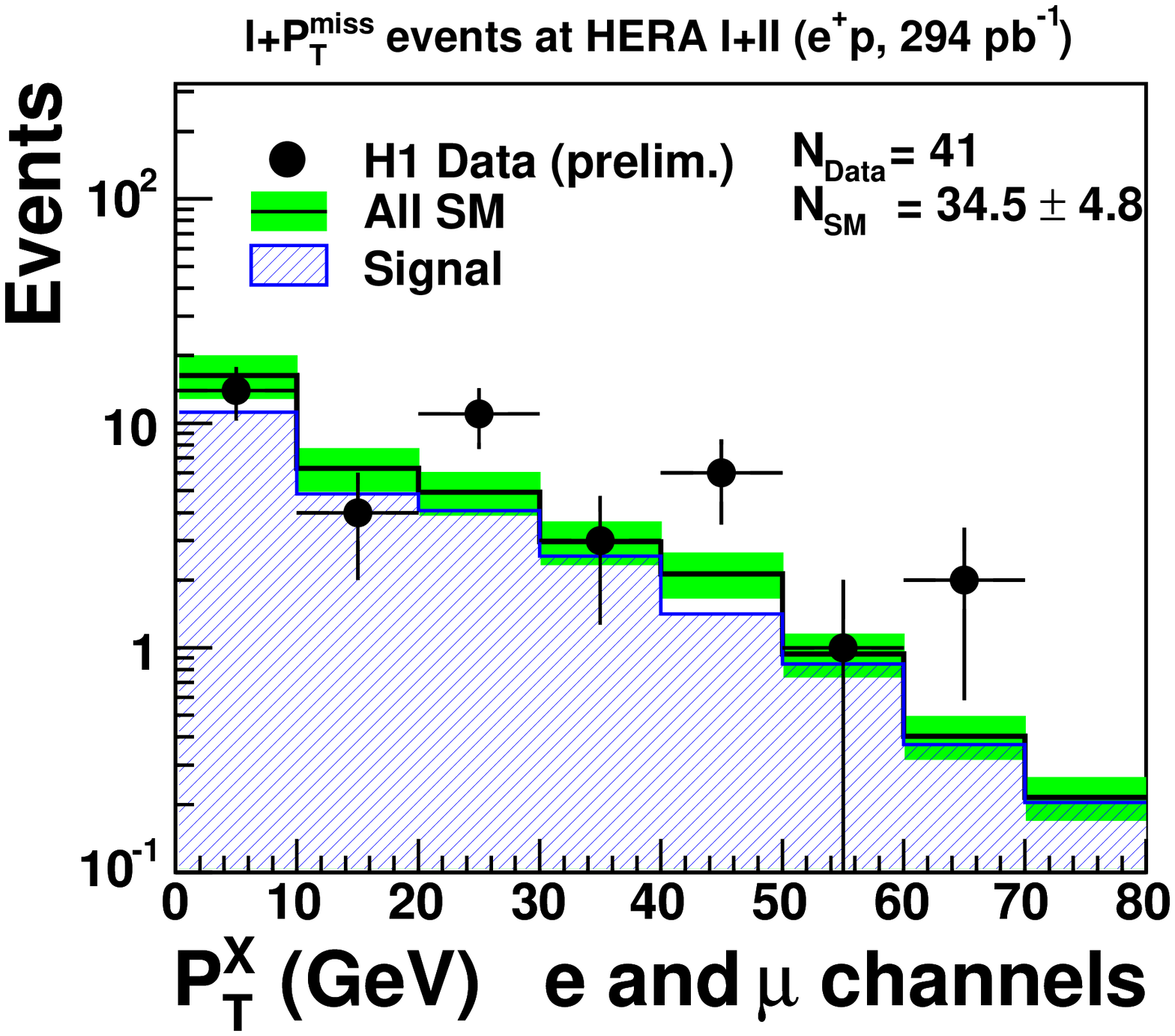}
  \end{center}
  \vspace{-0.75cm}
  \caption{Distributions of the transverse momentum of the hadronic system $P_T^X$ for the electron and muon channels combined in the $e^{-}p$~(left) and $e^{+}p$~(right) data.}
\label{fig:IsolatedLepton}  
\end{figure}

Events containing an isolated electron or muon, large missing transverse momentum and an high transverse momentum ($P_T^X$) hadronic system have been investigated by the H1 Collaboration~\cite{h1EPS07}. The full HERA data corresponding to a luminosity of $478$~pb$^{-1}$ are used. The main SM contribution to such a signature comes from the production of a real $W$ boson with subsequent leptonic decay. Figure~\ref{fig:IsolatedLepton} shows the observed $P_T^X$ distributions separately for the $e^{+}p$ and $e^{-}p$ data together with the corresponding SM expectations. For $P_T^X$~$>25$~GeV, $24$ data events are observed compared to a SM prediction of $15.8~{\pm}~2.5$. In this region, an excess of $e^{+}p$ data events is observed, equivalent to a fluctuation of order $"3\sigma"$. The observation in the $e^{-}p$ data is consistent with the SM expectation.
\vspace{-0.35cm}

\begin{figure}[htbp] 
  \begin{center}
    \includegraphics[width=5.2cm]{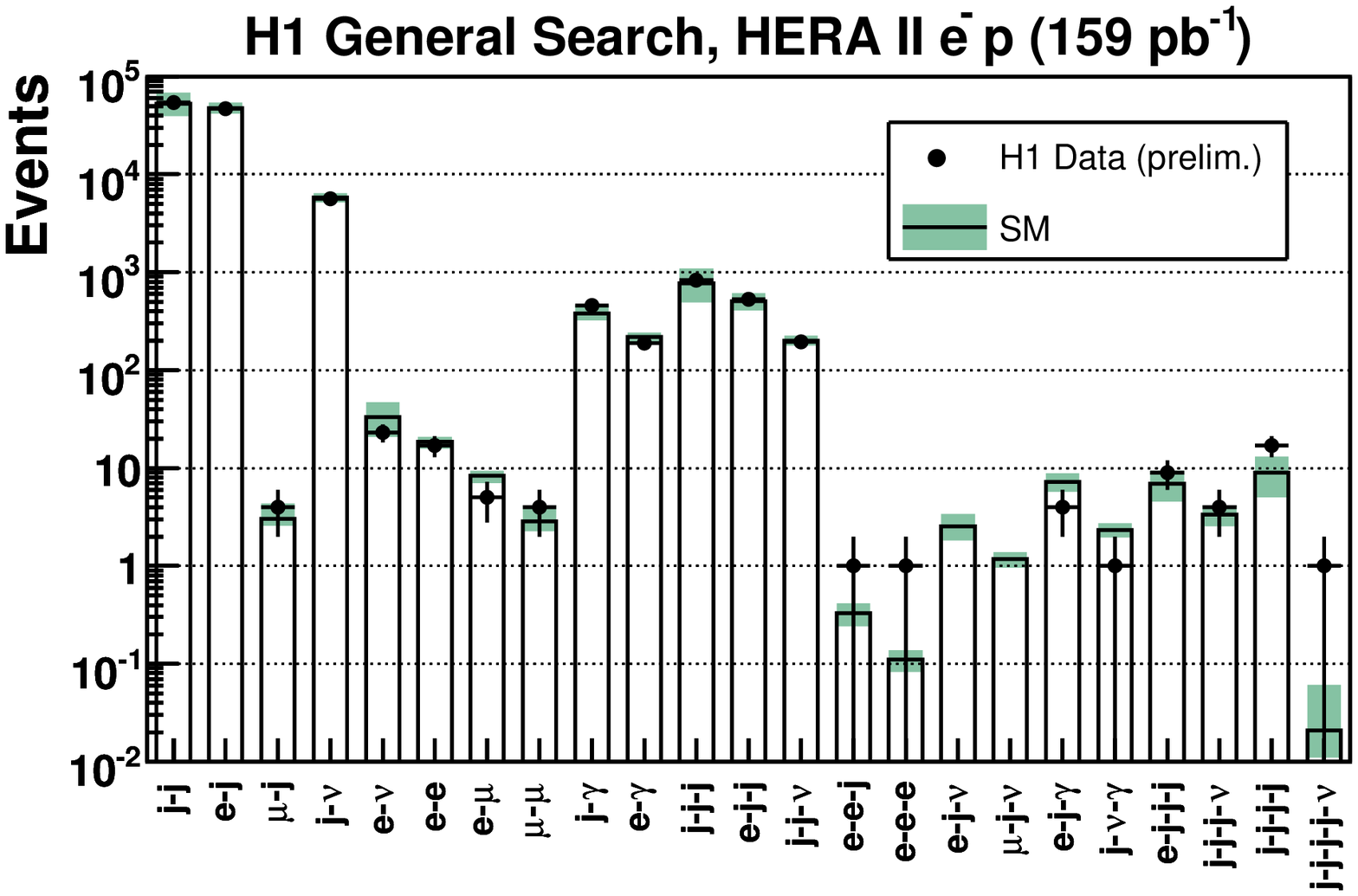}
    \includegraphics[width=5.2cm]{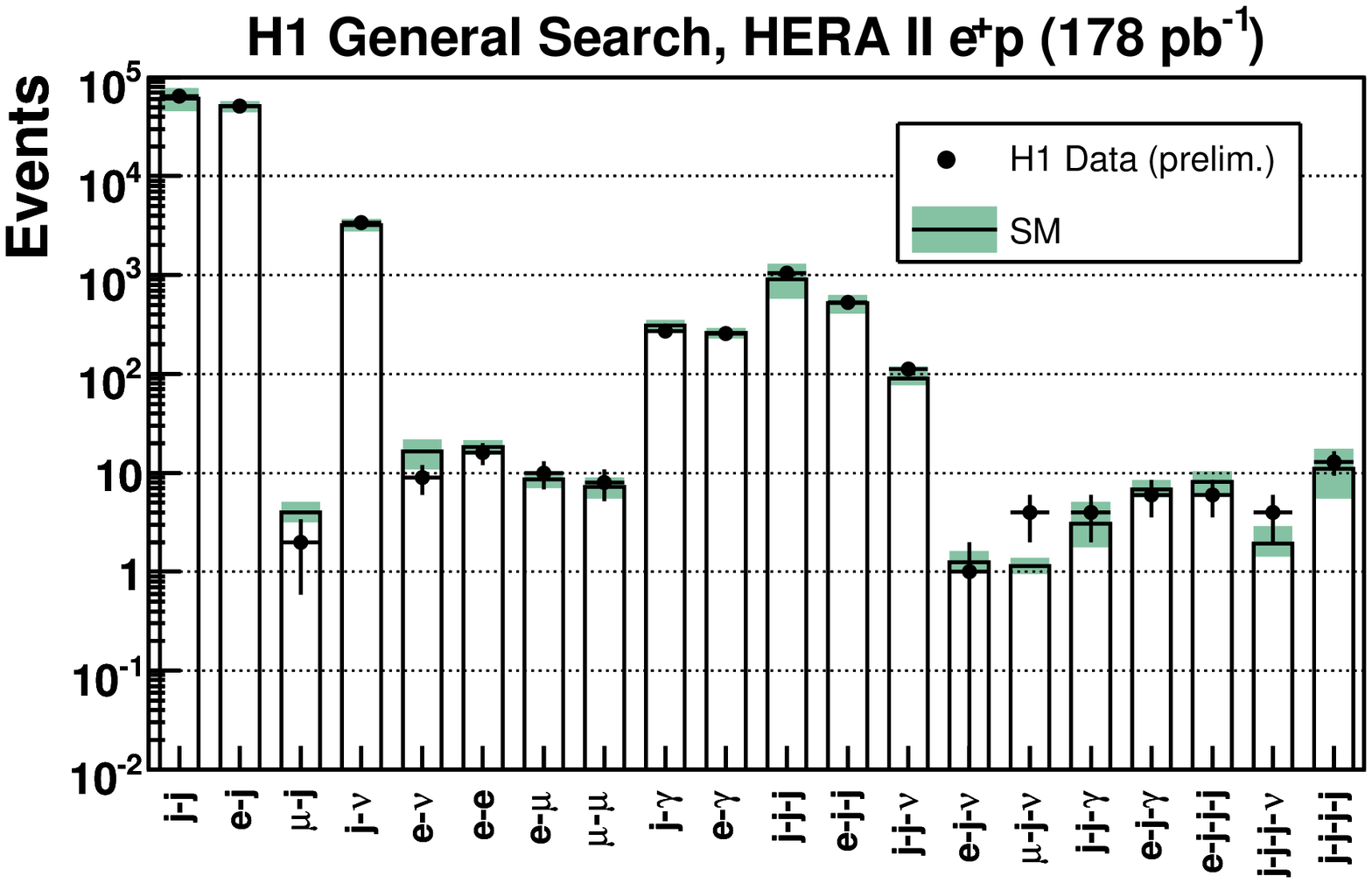}
  \end{center}
  \vspace{-0.7cm}
  \caption{The data and the SM expectation for all event classes with observed data events or a SM expectation greater than 1 event.}
\label{fig:GeneralSearch}  
\end{figure}

A model independent general search for deviations from the SM has been performed by H1 using all HERA II data. Following~\cite{GeneralHera1}, all final states containing at least two objects ($e,\mu,j,\gamma,\nu$) with $P_T>20$~GeV in the central region of the detector are investigated. Figure~\ref{fig:GeneralSearch} shows the event yield subdivided into event classes for the data and SM expectation for $e^+p$ and $e^-p$, respectively. A general good agreement is observed between the data and the SM prediction. A statistical algorithm is used to quantify the significance of possible deviations from the SM. The largest deviation is observed in $e^+p$ data in the ${\mu}$-$j$-$\nu$ channel, which corresponds to the topology of isolated lepton events.
\vspace{-0.35cm}


\vspace{-0.75cm}
\begin{thebibliography}{9} 
\vspace{-0.35cm}
\bibitem{Harari}
H.~Harari, Phys. Rept. {\bf 104} (1984) 159.
\bibitem{Boudjema}
F.~Boudjema, A.~Djouadi and J.~L.~Kneur, Z. Phys. {\bf C 57} (1993) 425
\bibitem{h1EPS07}
H1 Collaboration, contributed paper to {\it The International Europhysics Conference on High Energy Physics, EPS07}, Manchester 2007, {\bf Abstract 228}.
\bibitem{GeneralHera1}
H1 Collaboration, A.~Aktas et~{\it al.}, Phys.\ Lett.\ {\bf B 602} (2004) 14.
\end{thebibliography}
\end{document}